# Bayesian state estimation unlocks real-time control in thin film synthesis

*Sumner B. Harris[1]\*, Ruth Fajardo[2], Alexander A. Puretzky[1], Kai Xiao[1], Feng Bao[2], Rama K. Vasudevan[1]\**

1. Center for Nanophase Materials Sciences, Oak Ridge National Laboratory, Oak Ridge, Tennessee 37831, United States.
2. Department of Mathematics, Florida State University, Tallahassee, Florida 32306, United States.

   *Correspondence should be addressed to: harrissb@ornl.gov or vasudevanrk@ornl.gov

**Abstract**

The rapid validation of newly predicted materials through autonomous synthesis requires real-time adaptive control methods that exploit physics knowledge, a capability that is lacking in most systems. Here, we demonstrate an approach to enable the real-time control of thin film synthesis by combining *in situ* optical diagnostics with a Bayesian state estimation method. We developed a physical model for film growth and applied the Direct Filter (DF) method for real-time estimation of nucleation and growth rates during pulsed laser deposition (PLD) of transition metal dichalcogenides. We validated the approach on simulated and previously acquired reflectivity data for WSe$_2$ growth and ultimately deployed the algorithm on an autonomous PLD system during growth of 1T'-MoTe$_2$ under various synthesis conditions. We found that the DF robustly estimates growth parameters in real-time at early stages of growth, down to 15% monolayer area coverage. This approach opens new opportunities for adaptive film growth control based on a fusion of *in situ* diagnostics, modern data assimilation methods, and physical models which promises to enable control of synthesis trajectories towards desired material states.

# Main

The last decade has seen substantial investment into machine learning and high-throughput computational methodologies for materials science, in an endeavor to identify new materials with desirable properties. This is exemplified by the Materials Genome Initiative[1] which has paved the way for high-throughput computational screening, and driven research in automated experimentation. The vast parameter space of chemical structures can be efficiently navigated *in silico* via high-throughput density functional theory or molecular dynamics. The results of these efforts are stored in databases such as the Materials Project[2], which machine learning models can utilize to predict candidates for experimental synthesis of new materials.

Despite years of effort, the key obstacle for realizing the potential of such workflows remains the same: predicting new materials is straightforward compared to experimental validation through some synthesis modality. Recent advances in autonomous experiments have increased throughput for synthesis of thin films[3-5], nanoparticles[6], and single crystals[7] but very few efforts have successfully incorporated theory or literature data to narrow the parameter space and inform the commonly used Bayesian optimization loop with physics knowledge. Moreover, many predicted materials are metastable[8] which requires that synthesis trajectories be carefully controlled to drive the system towards the desired state. Controlling growth requires an ability to exploit *in situ* diagnostics collected in real-time to inform rapid changes to the trajectory of growth based on available data, models, and inferences of the future state of the system.

Model-based predictive control[9] (MPC) is a method that explicitly uses a process model to predict the future behavior of a system in real-time and has been widely used in chemical synthesis and process control, dating back to the 1970s[10,11]. Traditional feedback control methods such as proportional-integral-derivative (PID) loops are used to maintain system parameters at a setpoint but are purely reactive in nature which precludes their use for synthesis guided by a physical model. MPC predicts the state of the system at future times, allowing for adjustment of the input parameters based on a fusion of model predictions and current measurements, enabling manipulation of the system's trajectory. Implementing MPC for common materials synthesis methods would present a significant advance in the field.

MPC methods in thin film synthesis with physical vapor deposition (PVD) techniques are rarely applied due to challenges in deriving physical quantities from diagnostic data in real-time and the lack of compatible film growth models. For example, reflection high energy electron diffraction (RHEED) is ubiquitous in molecular beam epitaxy (MBE) and pulsed laser deposition (PLD) to monitor effective growth rates[12]. First MPC-like efforts in PVD date back to 1984 with "phase-locked epitaxy" where the MBE source shutter is controlled based on oscillations in the RHEED signal to terminate growth[13]. However, modeling realistic RHEED images from surface structures is an active research area, leading sporadic efforts to focus on neural network models for prediction and control. Neural networks for MBE control date back to 1993, again for shutter control to reliably achieve desired film thickness[14]. Later, neural networks were explored to analyze RHEED data in MBE to make a predictive model of the future diffraction state[15]. Modern work focuses on RHEED pattern classification and clustering implemented mostly post-growth[16-18]. While effective, these approaches still lack interpretability and physics awareness, which are crucial for guiding systems toward desired states.

To realize physics informed control over thin film synthesis, different methods are needed to infer the future state of the system. Optical diagnostics can be used with simple and accurate physics models, in contrast to RHEED, and provide similar information related to growth rates and composition. *In situ* ellipsometry has been used in an MPC scheme to modulate the composition of $Si_{1-x}Ge_x$ films during chemical vapor deposition (CVD) [19]. *In situ* Raman has also been shown to be sensitive to composition and strain during CVD[20] and PLD[21] but has not yet been used for MPC. Images of the plasma plume dynamics in PLD have been linked to film growth kinetics[22], but the deep learning model used does not provide interpretable correlations between the plume and kinetics.

Here, we combine *in situ* optical diagnostics with a recently developed Bayesian state estimation method towards enabling MPC of thin film nucleation and growth rates during PLD. We developed a simple and interpretable physical model for thin film growth based on the area coverage of discrete layers, which is measurable by laser reflectivity, and apply the direct filter (DF) method of parameter estimation to determine the growth and nucleation rates of thin films in real-time during PLD. We deployed our method on an autonomous PLD system and tested the algorithm in real-time during growth of 1T'-$MoTe_2$. We show that the DF can accurately estimate the growth model parameters at early stages of synthesis and is able to determine the nucleation and growth rates of the first monolayer after only ~15% area coverage has been deposited. We posit that this approach gives access to fundamental film growth kinetics early enough in the growth process to inform decisions to adaptively alter the trajectory of synthesis, possibly towards predicted metastable states. Any physical model describing film growth in the context of *in situ* diagnostics can be adapted to our method, providing a powerful tool to advance thin film synthesis by integrating modern data assimilation techniques with these diagnostics and physical models.

**Modeling Growth for In Situ Monitoring with Reflectivity**

For the growth of transition metal dichalcogenide (TMD) thin films with PLD, previous work showed that *in situ* laser reflectivity reveals sub-monolayer growth and nucleation kinetics on $SiO_2$/Si substrates, where the multilayer structure enhances reflected contrast[23]. Building on this, we developed a general growth model for arbitrary number of TMD layers and calculated reflected contrast using the Fresnel equations and fractional area coverage of individual layers. We use the same recursive method for calculating the Fresnel reflection coefficient of a homogeneous layer stack and refer to the previous work for a detailed description[23]. Thus, our model describes the time-evolution of the fractional area coverage for discrete single-layers to simulate the experimentally observed reflected contrast.

To describe growth kinetics of 2D materials, we employ a two-step kinetic model that considers conversion of A to B through the nucleation and autocatalytic growth steps described by the rate constants, $k_n$ and $k_{gr}$, respectively.

$$A \xrightarrow{k_n} B, \text{nucleation} \quad \text{(step 1)}$$

$$A + B \xrightarrow{k_{gr}} 2B, \text{growth} \quad \text{(step 2)}$$

This approach is the most general and can be applied to interpret sigmoidal kinetics found in many different processes that exhibit cooperative effects when the initial conversion of A to B affects the subsequent conversion process. As shown by Finney et al.[24], this approach can be used to interpret the parameters of the Avrami equation[25-27] that was initially developed in the 1940's for kinetics of phase changes as well as its numerous modifications and derivatives (see Finney et al.[24] for review), e.g., for kinetics of diamond films deposition by CVD[28]. Here, we apply this growth model to PLD growth of TMDs.

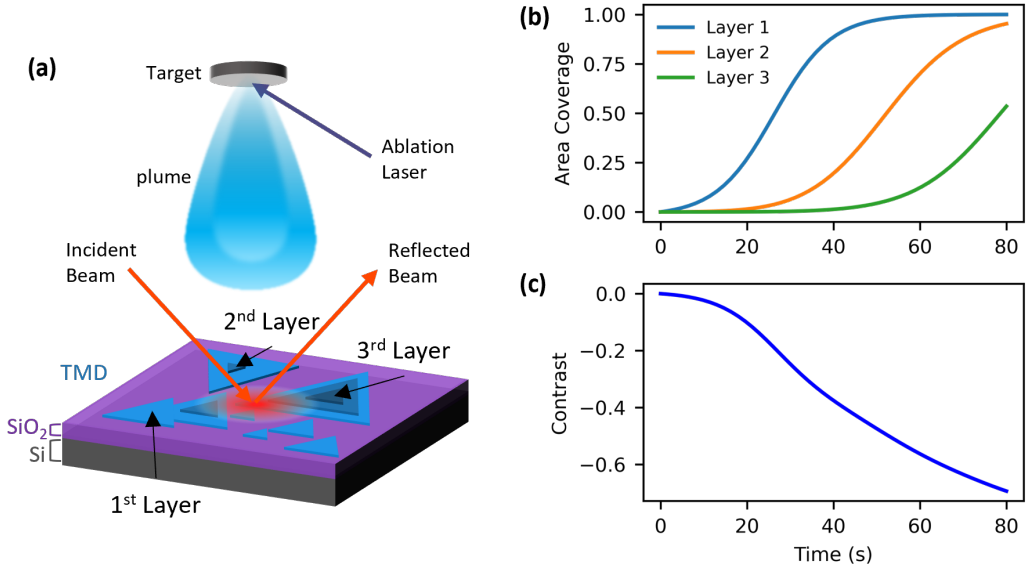

**Figure 1**. a) Schematic of pulsed laser deposition of few-layer transition metal dichalcogenides (TMDs) with *in situ* laser reflectivity to monitor growth kinetics. Film growth can be described in terms of fractional area coverage of discrete layers which nucleate and grow at different rates during deposition. b) An example of the area coverage vs. time for 3 layers of TMD growth based on our autocatalytic growth model. c) The calculated reflected contrast based on the layer coverage shown in b) vs time.

**Figure 1a** shows a schematic of the experimental arrangement for PLD synthesis and the layer stack model used to calculate the contrast. For TMD synthesis with PLD, a solid target of the desired TMD material is irradiated with a pulsed laser to create a plasma plume which contains the vapor phase precursors for film growth. The plume species condense on the SiO$_2$/Si substrate where monolayer islands begin to nucleate with a rate $k_{n1}$. Other particles diffuse along the substrate surface and attach to existing islands with a growth rate of $k_{gr1}$. The initial area available for a monolayer growth is $f_0(t=0) = 1$ with the initial value of the fractional area coverage of the first monolayer $f_1(t=0) = 0$. At some point, additional layers begin to nucleate and grow on top of the 1$^{st}$ layer islands, and so on. This nucleation and growth process can be generalized by a system of equations given in Eq. 1, where $f_i$ is the fractional area coverage of layer number *i* with nucleation and growth rates of $k_{ni}$ and $k_{gri}$, respectively, up to a total of N layers. Note that here $k_{gri}$ is multiplied by the initial surface area, $A_0 = 1$, to remove the area dependence and to make both rate dimensions s$^{-1}$. The reflected contrast is given by $C_i = (R_i - R_0)/R_0$ where $R_i$ is the Fresnel reflection coefficient of a layer stack with $i$ TMD layers and $R_0$ is the reflection coefficient of the bare substrate. Thus, we can model the time-evolution of the contrast $C_r(t)$ with Eq. 2.

$$\frac{df_i}{dt} = k_{ni}(f_{i-1} - f_i) + k_{gri}(f_{i-1} - f_i)f_i \text{ where } i = 1,2,\dots,N \tag{1}$$

$$C_r(t) = \sum_{i=1}^{N}(C_i - C_{i-1})f_i(t) \tag{2}$$

**Figure 1b** shows the individual simulated area coverage vs time for the first 3 TMD layers while **Figure 1c** shows the corresponding overall contrast $C_r(t)$. Eq. 1 can be solved numerically for a defined number of layers. For simplified case studies of a single monolayer growth, we can solve Eq. 1 analytically (Eq. 3) where $f$, $k_n$, and $k_{gr}$ are the first monolayer coverage, and nucleation and growth coefficients, respectively.

$$f(t) = \frac{\frac{k_n}{k_{gr}}\left(e^{(k_n + k_{gr})t} - 1\right)}{1 + \frac{k_n}{k_{gr}}e^{(k_n + k_{gr})t}} \tag{3}$$

This growth model explains the experimentally observed *in situ* reflectivity and allows for on-line or off-line fitting methods to estimate the fundamental nucleation and growth rates for individual TMD layers during PLD synthesis. Specifically for this work, we aim to apply this model with on-line, Bayesian methods to estimate the nucleation and growth rates with uncertainty to enable real-time monitoring of these quantities during PLD.

**Direct Filter Method for Parameter Estimation**

The ability to accurately predict fundamental parameters related to thin film growth in real-time is highly attractive towards precisely controlling synthesis. Cast in terms of a state estimation problem, Eq. 1-2 represent the state-space model and the observation model for film growth where $k_{ni}$ and $k_{gri}$ are the model parameters. Here, we are interested in estimating the parameters of the state space-model, rather than the state itself (the contrast), in real-time. To do this, we apply the direct filter (DF) method[29] which is a nonlinear particle filtering method that accurately estimates dynamical state-space model parameters.

Filtering methods in data assimilation fuse information from both physical model simulations and observed data within specific time frames. This fusion aims to refine our understanding of a dynamical system and its associated uncertainties recursively in time. Suppose one has the following state-space model $X_{n+1} = G(X_n, \theta) + w_n$, $n = 1,2,3,\cdots$, where $G(X_n, \theta)$ is the dynamical model describing the evolution of the state process $X_n$ at time step $n$ defined by the state parameters $\theta$, where $w_n$ represents additive noise that perturbs the system. Alongside, we have an observation model $Y_{n+1} = H(X_{n+1}) + \xi_{n+1}$, where $H$ represents the observation function and $\xi_{n+1}$ is the observation noise. The DF method considers the model parameters $\theta$ as the only state to estimate. Since the observation process does not directly measure the parameters, the DF composites the state model into the observation function to construct the nonlinear filtering problem given by Eq. 4a-b where $\epsilon_n$ is additive Gaussian noise.

$$\theta_{n+1} = \theta_n + \epsilon_n \tag{4a}$$

$$Y_{n+1} = H(G(X_n, \theta_{n+1}) + w_n) + \xi_{n+1} \tag{4b}$$

A detailed description of the DF implementation used in this work is given in supplemental information (**Note S1**). We first apply DF estimation to synthetic data generated using the monolayer model (Eq. 3) to simultaneously estimate the nucleation and growth rates $k_n$ and $k_{gr}$ of WSe$_2$ monolayers to evaluate the efficacy of this approach. We find that the DF can quickly and accurately estimate the rate parameters when using a "burn-in" period that set the artificial noise $\epsilon_n = 0$ after a short time. Full details of this synthetic data study are given in supplemental information (**Note S2**, **Figure S1**).

**Sequential Estimation of Monolayer Growth**

We test the DF method with previously acquired experimental data that was collected during an autonomous WSe$_2$ growth experiment[5]. **Figure 2a-c** shows the sequential estimation of $k_n$ and $k_{gr}$ along with the predicted contrast, respectively, for experimental data. We repeated the estimation 10 times with different a random seed to check for repeatability using a burn in time of 8s. We observe similar convergent behavior as in the synthetic data case (**Figure S1**) and find the 10-trial average values of $k_n = (2.9 \pm 0.3) \times 10^{-3}$ s$^{-1}$ and $k_{gr} = 0.147 \pm .009$ s$^{-1}$. This method enables projection of the contrast curve to future times, which is one possible application during on-line estimation. We take the parameter estimates at fixed times after the burn-in period and project the contrast forward with uncertainty. **Figure 2d** shows the results of these projections at 8.3 s, 11.3 s, 15.3 s, and 16.9 s. The uncertainty is high immediately after burn-in but decreases as more data is collected. The projections at 15.3 s and 16.9 s (approximately twice the burn-in time) both have a small enough uncertainty that they could be used to reasonably predict the final growth time and rates with enough remaining time to make changes to the experimental parameters and alter the course of growth. The contrast at these two times corresponds to 15% and 19% monolayer area coverage and correspond to a difference of only 2-3 laser pulses. At these low coverages and at the lower limit of deposition control via laser pulse number, we anticipate that intervening in the experiment at this time by changing the repetition rate, laser fluence, or substrate temperature could meaningfully alter the future growth and resulting properties of the film. This capability is highly desirable for real-time, adaptive control of thin film growth with PLD.

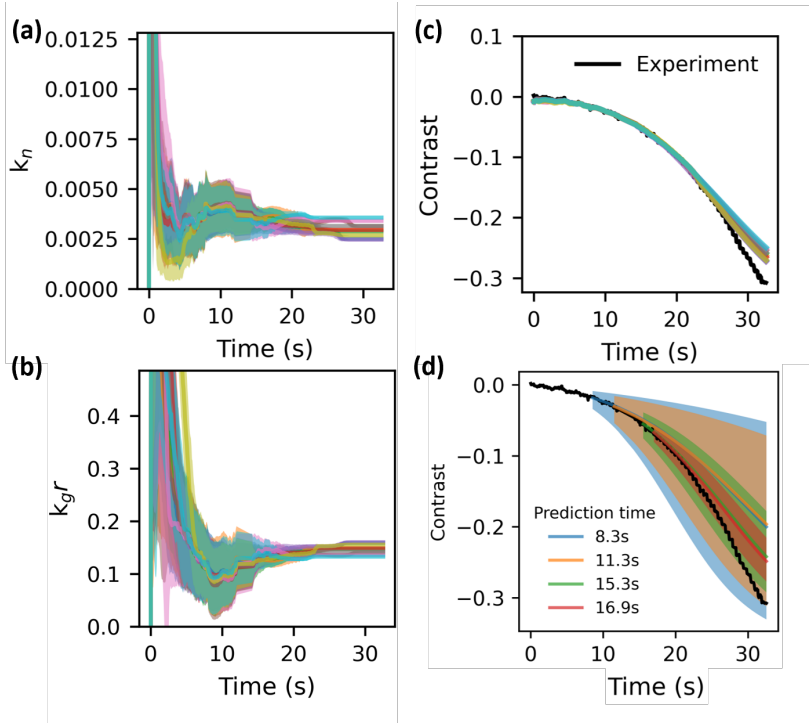

**Figure 2**. Sequential parameter estimation using experimental data with the monolayer model (Eq. 3) for 10 trials with an 8 s burn-in period. Each trial is shown as a different color line where the shaded region represents the uncertainty. a) nucleation rate $k_n$ and b) growth rate $k_{gr}$ are estimated for each sequential timestep and c) shows the predicted contrast. The average final parameter estimates for this experiment are $k_n = (2.9 \pm 0.3) \times 10^{-3}$ and $k_{gr} = 0.147 \pm .009$. d) The projected contrast with uncertainty (shaded area) at four times after the burn-in period.

Notably, the predicted contrast deviates from the experimental measurements at later times (**Figure 2c**). This is due to the nucleation of additional layers beyond the first monolayer whose fractional area coverage begins to significantly contribute to the reflected contrast. We previously found that beyond ~40% monolayer coverage, the 2nd layer begins to nucleate[23]. Indeed, the contrast at which the experiment deviates from the monolayer growth model, at ~ 25 s, corresponds to 45% monolayer coverage so we can anticipate 2$^{nd}$ layer nucleation.

**Real-time Parameter Estimation during Automated PLD Growth**

To our knowledge, state estimation methods such as the one developed in this work have never been deployed on any PLD system. To demonstrate this application in a real PLD environment, we integrated the above monolayer growth model and DF algorithm into the control software for an autonomous PLD system[5] and grew ultrathin films of 1T'-MoTe$_2$ at different Ar background pressures to test the reliability of real-time parameter estimation under conditions with different deposition rates. We synthesized 2 films at 5 different pressures between 0-70 mTorr with all other deposition variables held constant and we also fixed the DF algorithm parameters, details are given in the Methods section. Deposition was automatically terminated when the contrast reached the value expected for a full coverage monolayer of 1T'-MoTe$_2$ which was -0.54.

MoTe$_2$ can crystallize in three different phases: the hexagonal semiconducting 2H phase, the monoclinic metallic 1T' phase, or the orthorhombic T$_d$ phase which exhibits quantum type-II Weyl semimetal behaviour[30] and superconductivity[31]. The 2H phase is the most thermodynamically stable, but the energy difference between the 2H and 1T' phases is the lowest of all TMDs at only ~35 meV[32]. The 1T' phase tends to form at higher growth temperatures[33] and under Te-deficient conditions[34] while the T$_d$ phase becomes the most stable at high temperatures[35] compared to the 2H phase. Therefore, MoTe$_2$ is an ideal material system for exploring synthesis methodologies to tune the phase compositions of thin films. Although no direct growth of MoTe$_2$ by PLD has been reported, there are two studies on PLD deposited amorphous (Mo,W)Te$_{2-x}$ alloys at room temperature with post-growth annealing[36,37].

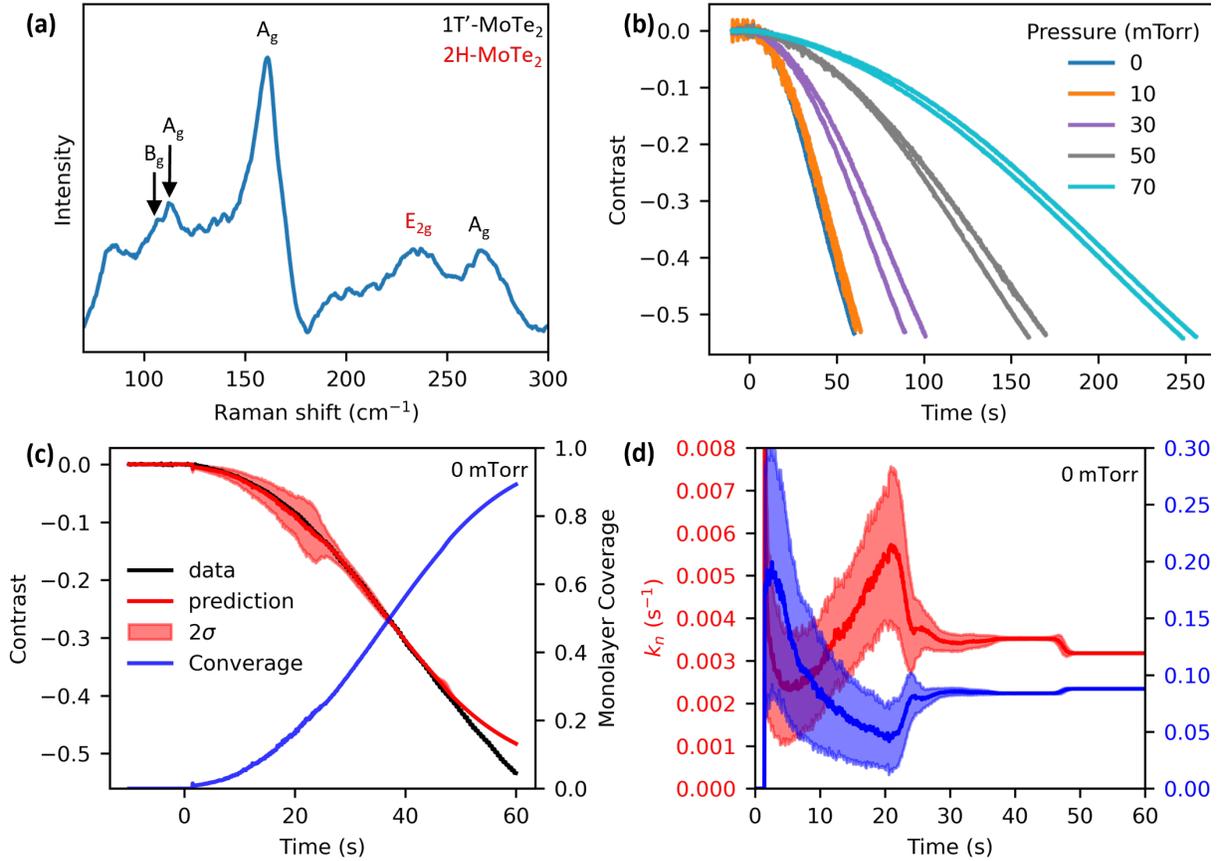

**Figure 3**. Application of real-time direct filter (DF) parameter estimation during PLD synthesis of ~ 1 monolayer thick 1T'-MoTe$_2$ films grown at 200 °C with various Ar background pressures between 0-70 mTorr. a) Raman spectrum of ~10 layer thick film indicates the predominant phase is the metallic 1T' with some semiconducting 2H present. b) Laser reflectivity contrast curves show the difference in growth rates at different pressures, with some run-to-run variability. c) Detailed view of contrast curve for the film grown at 0 mTorr comparing the experimental data (black), with the DF predictions (red), and the monolayer area coverage (blue). The envelope around the DF predictions is the uncertainty (+/- std. dev., σ). d) DF predicted growth and nucleation rates, $k_n$ and $k_{gr}$, with uncertainty vs. time.

**Figure 3a** shows the Raman spectrum of a typical thick film of MoTe$_2$ grown at 200 °C (~10 layers, determined by reflected contrast). The spectrum indicates that the films crystallize predominantly in the 1T' phase, with the main $A_g$ modes of 1T'-MoTe$_2$ at 161 cm$^{-1}$ and 266 cm$^{-1}$, and other broad peaks in the vicinity of the $B_g$ mode at 107 cm$^{-1}$ and $A_g$ mode at 110 cm$^{-1}$, consistent with literature[38,39]. The weaker Raman peaks are not resolvable, so we only refer to the general regions expected for these modes. The spectrum we obtain is similar to nanostructured 1T' films[40] and also of the Te-deficient films in Sun et al.[36] Interestingly, the 2H phase should be more stable than the 1T' phase below ~300 °C[41]. The 1T' phase is the dominant phase in our case due to Te deficiency, most likely caused by incommensurate evaporation of Mo and Te during ablation. It has been shown that excess Te is required to stabilize the 2H phase and that the 1T' phase forms otherwise[34,42]. Film optimization and properties are not the topic of this work and so will be left to future studies. Here, knowledge of the primary crystal phase is required to select the correct optical constants to conduct the DF experiments. We selected the refractive index of 1T'-MoTe$_2$ in this case, given in the Methods section.

**Figure 3b** shows the real-time reflected contrast during each growth. The number of pulses needed to grow ~1 monolayer ranges from 120 at 0 mTorr to 512 at 70 mTorr, with run-to-run variation likely caused by changes in the target surface that module the ablation yield with each laser pulse. This variation highlights the need for real-time control algorithms. DF parameter estimation was performed and recorded in real-time, without tuning the DF algorithm parameters during the synthesis of all 10 samples, demonstrating its robustness across varying growth rates. **Figure 3c** details one sample grown at 0 mTorr, where DF contrast predictions (red) with a 2σ uncertainty envelope (+/- 1 standard deviation σ) are compared to monolayer area coverage (blue). DF predictions deviate from the experimental data at ~ 49.5 s, aligning with ~76% monolayer coverage, beyond which we expect significant contributions from additional layers that the model does not consider. **Figure 3d** shows that parameter predictions stabilize after ~19.5 s, with final predicted rates of kn = $3.18 \times 10^{-3}$ s$^{-1}$ and kgr = $8.78 \times 10^{-2}$ s$^{-1}$. The smaller kgr/kn ratio of 27.6 for 1T'-MoTe2 compared to WSe2 (50.7) can explain the higher monolayer coverage (76% vs. 40%) before additional layer nucleation.

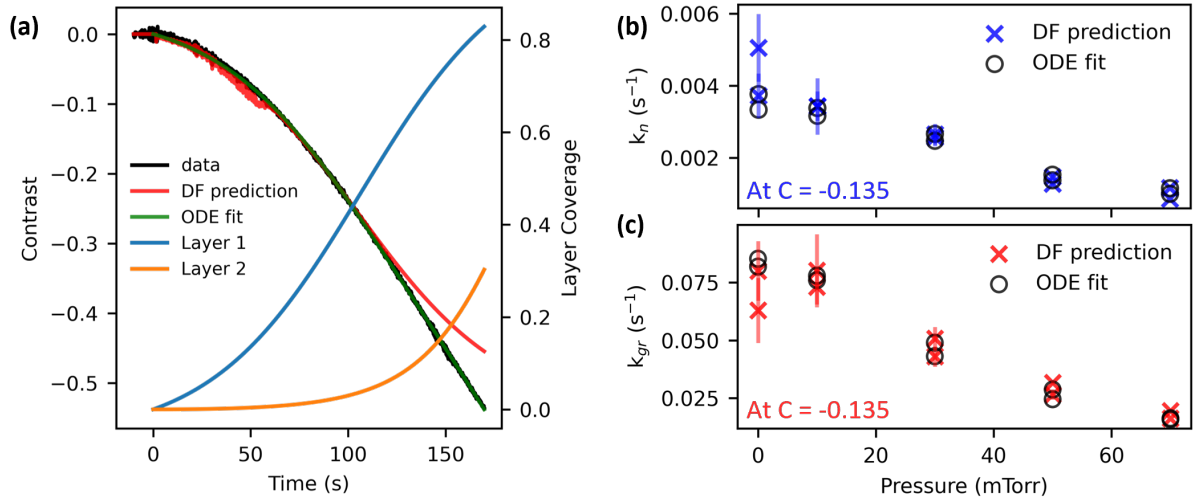

**Figure 4**. Comparison of real-time direct filter (DF) parameter estimation with post-growth fitting with a 2 layer model. a) Contrast curve during PLD growth of $MoTe_2$ at 50 mTorr overlaid with the real-time DF predictions and the post growth 2 layer model fit (ODE fit) (left axis). The individual layer coverages from the ODE fit (right axis) show that the monolayer DF model begins to diverge from the data as the layer 2 begins to grow, as expected. The nucleation rate $k_n$ b) and growth rate $k_{gr}$ c) for layer 1 vs. Ar pressure determined from the DF and ODE fits closely match. The ODE fit requires all the data (post growth analysis) while the rates shown in b-c) for the DF were taken from the point of each curve when the contrast reached -0.135 (~25% monolayer coverage). The DF parameter estimation can capture the growth kinetic parameters accurately in real time, at early stages of film growth.

Finally, we compared real-time DF rate predictions with post-growth analysis to determine if the rates are accurately predicted prior to the end of film growth. Using a two-layer model (Eq. 1-2), we fit all 10 $MoTe_2$ contrast curves, solving the equations via integration (SciPy[43] odeint) and minimizing MSE as a function of $k_{ni}$ and $k_{gri}$ (ODE fit). **Figure 4a** shows one 50 mTorr 1T'-MoTe2 deposition, with real-time DF predictions overlaid with ODE fit results. Both methods fit the data well, though the DF deviates when the second layer forms, expected as previously noted. The key difference is that the DF is done is real-time as data is received (partial data) from the detector whereas the ODE fit requires the whole curve. **Figure 4b,c** display $k_n$ and $k_{gr}$ vs Ar pressure for layer 1 using DF estimation at ~25% monolayer coverage (contrast of -0.135 ) for each deposition compared with ODE fit results. DF accurately matches the post-growth ODE fits, even with partial data early in deposition. Thus, this method is highly effective for tracking growth kinetics in real-time, supporting future autonomous optimization during thin film synthesis via PLD.

## Conclusions

We applied an on-line Bayesian state estimation technique called the direct filter method (DF) to estimate the parameters of a thin film growth kinetics model in real-time during pulsed laser deposition (PLD) of ultrathin transition metal dichalcogenide (TMD) materials. The DF estimates growth and nucleation rate parameters of a growth model based on area coverage of discrete layers, measurable with *in situ* laser reflectivity. We tested the method on synthetic and previously acquired data for $WSe_2$ growth and ultimately deployed the algorithm on an autonomous PLD system to demonstrate real-time parameter estimation during automated growth of 1T'-$MoTe_2$ under different conditions. The DF method robustly estimates model parameters at early stages of growth with accuracy consistent with post-growth analysis.

The DF approach highlights the utility of exploiting recent developments in applied mathematics towards control problems in materials synthesis, addressing a key bottleneck in materials discovery: the inability to inform synthesis processes with physical models in real time. For deposition methods such as PLD, real-time control is almost nonexistent due to the absence of viable physical models for parameter inference, making model-based control impractical. By combining real-time diagnostics with a physical model, real-time control in PLD becomes feasible. Extending this approach to more complex growth models (e.g., kinetic monte-carlo), additional *in situ* data (e.g., RHEED, Auger electron spectroscopy, etc.), and advanced control methods like reinforcement learning has promise to realize true control of synthesis for thin films and may finally enable tailored synthesis of desired metastable phases that are very difficult or impossible to reliably fabricate using existing methodologies.

**Methods**

Pulsed laser deposition of 1T'-MoTe$_2$ films was conducted by ablating a MoTe$_2$ target (99.8%, Plasmaterials, Inc.) using a KrF excimer laser (Coherent LPX 305F, 248 nm, 25 ns pulse width, 2 Hz repetition rate) with a target to substrate distance of 5 cm. The target was offset from the substrate by 25°. The KrF beam was passed through a 10×10 mm$^2$ aperture which was imaged onto the target using a projection beamline to produce a square spot with an area of 0.0256 cm$^2$. The laser energy was 25.6 mJ for a fluence of 1.0 J/cm$^2$. The Ar background pressure (99.9999%) was regulated by a throttle valve and a mass flow controller, using 0.5 sccm for 10 mTorr, and 5 sccm all other pressures. Substrates were heated from the backside with a remote 976 nm, 140 W laser directed at Inconel sample holders and the temperature was measured a pyrometer to within ±1 °C. The substrate temperature for all depositions was 200 °C. The 5 × 5 mm$^2$ substrates used for growth were all diced from the same 3-inch 90 nm SiO$_2$/Si wafer (University Wafer, ID: 3595, Dry Thermal Oxide). The substrates were prepared by sonication in acetone, methanol, and isopropyl alcohol, followed by blow drying with nitrogen. Silver paste was used to bond the substrates to the Inconel sample plates and were baked on a hot plate for 20 min at 130 °C before loading into the chamber. The base pressure achieved before deposition was less than 1×10$^{-6}$ Torr.

During automated PLD growth of 1T'-MoTe$_2$, the DF algorithm parameters were held fixed. All runs used the same initial guess of $\theta_0$ = (0.002 s$^{-1}$, 0.01 s$^{-1}$) and artificial noise variance $\Sigma$ = (1.0×10$^{-6}$ s$^{-2}$ , 1.0×10$^{-3}$ s$^{-2}$) for $k_n$ and $k_{gr}$, respectively, and 3000 particles. Because films grown at different pressures will have different characteristic nucleation and growth time scales, we used different burn-in times for each pressure that were automatically determined in real-time based on the contrast value. The burn-in period for each film was maintained until the measured contrast reached -0.08 (corresponding to ~ 15% monolayer area coverage), after which $\epsilon_n$ was set to 0, which is a more flexible approach than a fixed time period.

Reflectivity was monitored using a stabilized HeNe laser (632.8 nm, 1.2 mW, Thorlabs, Inc, HRS015B) with an incident angle of 32.5°. The beam was randomly polarized using a liquid crystal polymer depolarizer (Thorlabs, Inc., DPP25-B). Reflected intensity was measured through a laser line filter (Thorlabs, Inc. FL632.8-1) using a photodiode (Thorlabs, Inc., SM1PD1B) and a source measure unit (Keithley 2450 SMU), streaming approximately 15 points per second to the parameter estimation algorithm.

Raman spectroscopy was done with a custom-built spectroscopy microscope using a 100× objective and 320 µW, 532 nm laser with an 1800 grooves/mm grating. Raman acquisition was 15 s exposure with 4 averages.

For modeling the Fresnel reflection coefficients, refractive indices of 3.87-0.016i and 1.47 were used for Si and SiO$_2$[44], 4.42-0.60i was used for WSe$_2$[45], and 4.19-2.03i for 1T'-MoTe$_2$[46].

## Conflict of Interest

The authors declare no competing interests.

## Author Contributions

**S.B.H.**: Conceptualization (lead); Investigation (equal); data curation (lead); methodology (equal); software (lead); visualization (lead)writing—original draft (lead); writing—review & editing (lead). **R.F.**: Methodology (equal); Investigation (equal); software (supporting); writing—original draft (supporting); writing—review & editing (supporting). **A.A.P**: Methodology (equal); Investigation (equal); writing—original draft (supporting). **K.X.**: writing—review & editing (supporting) **F.B.**: Methodology (equal); writing—review & editing (supporting). **R.K.V.**: writing—original draft (supporting); Conceptualization (supporting); writing—review & editing (supporting). All authors read and approved the final manuscript.

## Data Availability

The data that support the findings of this study are openly available at https://github.com/sumner-harris/PLD-Direct-Filter.git

## Code Availability

The data that support the findings of this study are openly available at https://github.com/sumner-harris/PLD-Direct-Filter.git

## Acknowledgements

This work was supported by the Center for Nanophase Materials Sciences (CNMS), which is a US Department of Energy, Office of Science User Facility at Oak Ridge National Laboratory. Materials synthesis was supported by the U.S. Department of Energy, Office of Science, Basic Energy Sciences, Materials Sciences and Engineering Division. FB would like to acknowledge the support from U.S. National Science Foundation through project DMS-2142672 and the support from the U.S. Department of Energy, Office of Science, Office of Advanced Scientific Computing Research, Applied Mathematics program under Grant DE-SC0025412.## References

20    Qin, Y. *et al.* Reaching the Excitonic Limit in 2D Janus Monolayers by In Situ Deterministic Growth. *Advanced Materials* **34**, 2106222 (2022). https://doi.org/https://doi.org/10.1002/adma.202106222

21    Harris, S. B. *et al.* Real-Time Diagnostics of 2D Crystal Transformations by Pulsed Laser Deposition: Controlled Synthesis of Janus WSSe Monolayers and Alloys. *ACS Nano* **17**, 2472-2486 (2023). https://doi.org/10.1021/acsnano.2c09952

22    Harris, S. B., Rouleau, C. M., Xiao, K. & Vasudevan, R. K. Deep learning with plasma plume image sequences for anomaly detection and prediction of growth kinetics during pulsed laser deposition. *npj Computational Materials* **10**, 105 (2024). https://doi.org/10.1038/s41524-024-01275-w

23    Puretzky, A. A. *et al.* In situ laser reflectivity to monitor and control the nucleation and growth of atomically thin 2D materials*. *2D Materials* **7**, 025048 (2020). https://doi.org/10.1088/2053-1583/ab7a72

24    Finney, E. E. & Finke, R. G. Is There a Minimal Chemical Mechanism Underlying Classical Avrami-Erofe'ev Treatments of Phase-Transformation Kinetic Data? *Chemistry of Materials* **21**, 4692-4705 (2009). https://doi.org/10.1021/cm9018716

25    Avrami, M. Kinetics of Phase Change. I General Theory. *The Journal of Chemical Physics* **7**, 1103-1112 (1939). https://doi.org/10.1063/1.1750380

26    Avrami, M. Kinetics of Phase Change. II Transformation‐Time Relations for Random Distribution of Nuclei. *The Journal of Chemical Physics* **8**, 212-224 (1940). https://doi.org/10.1063/1.1750631

27    Avrami, M. Granulation, Phase Change, and Microstructure Kinetics of Phase Change. III. *Journal of Chemical Physics* **9**, 177-184 (1941).

28    Tomellini, M. Coverage‐time dependence during island growth at a solid surface with application to diamond deposition from the gas phase. *Journal of Applied Physics* **72**, 1589-1594 (1992). https://doi.org/10.1063/1.351674

29    Archibald, R., Bao, F. & Tu, X. A direct filter method for parameter estimation. *Journal of Computational Physics* **398**, 108871 (2019). https://doi.org/https://doi.org/10.1016/j.jcp.2019.108871

30    Jiang, J. *et al.* Signature of type-II Weyl semimetal phase in $MoTe_2$. *Nature Communications* **8**, 13973 (2017). https://doi.org/10.1038/ncomms13973

31    Qi, Y. *et al.* Superconductivity in Weyl semimetal candidate $MoTe_2$. *Nature Communications* **7**, 11038 (2016). https://doi.org/10.1038/ncomms11038

32    He, H.-K. *et al.* Ultrafast and stable phase transition realized in $MoTe_2$-based memristive devices. *Materials Horizons* **9**, 1036-1044 (2022). https://doi.org/10.1039/D1MH01772A

33    Sung, J. H. *et al.* Coplanar semiconductor–metal circuitry defined on few-layer $MoTe_2$ via polymorphic heteroepitaxy. *Nature Nanotechnology* **12**, 1064-1070 (2017). https://doi.org/10.1038/nnano.2017.161

34    Park, J. C. *et al.* Phase-Engineered Synthesis of Centimeter-Scale 1T′- and 2H-Molybdenum Ditelluride Thin Films. *ACS Nano* **9**, 6548-6554 (2015). https://doi.org/10.1021/acsnano.5b02511

35    Ryu, H. *et al.* Anomalous Dimensionality-Driven Phase Transition of $MoTe_2$ in Van der Waals Heterostructure. *Advanced Functional Materials* **31**, 2107376 (2021). https://doi.org/https://doi.org/10.1002/adfm.202107376

36    Sun, W. *et al.* Sizable spin-to-charge conversion in PLD-grown amorphous $(Mo, W)Te_{2-x}$ films. *Nanotechnology* **34**, 135001 (2023). https://doi.org/10.1088/1361-6528/acaf34
21

# Supplemental Information for:

# Bayesian state estimation unlocks real-time control in thin film synthesis

*Sumner B. Harris[1]\*, Ruth Fajardo[2], Alexander A. Puretzky[1], Kai Xiao[1], Feng Bao[2], Rama K. Vasudevan[1]\**

1. Center for Nanophase Materials Sciences, Oak Ridge National Laboratory, Oak Ridge, Tennessee 37831, United States.
2. Department of Mathematics, Florida State University, Tallahassee, Florida 32306, United States.

\*Correspondence should be addressed to: harrissb@ornl.gov or vasudevanrk@ornl.gov

## Note S1: Direct Filter Method for Parameter Estimation

Suppose one has the following state-space model $X_{n+1} = G(X_n, \theta) + w_n$, $n = 1,2,3,\cdots$, where $G(X_{n,}\theta)$ is the dynamical model describing the evolution of the state process $X_n$ at time step $n$ defined by the state parameters $\theta$, where $w_n$ represents additive noise that perturbs the system. Alongside, we have an observation model $Y_{n+1} = H(X_{n+1}) + \xi_{n+1}$, where $H$ represents the observation function and $\xi_{n+1}$ is the observation noise. The DF method considers the model parameters $\theta$ as the only state to estimate. Since the observation process does not directly measure the parameters, the DF composites the state model into the observation function to construct the nonlinear filtering problem given by Eq. 1a-b where $\epsilon_n$ is additive Gaussian noise.

$$\theta_{n+1} = \theta_n + \epsilon_n \tag{1a}$$

$$Y_{n+1} = H(G(X_n, \theta_{n+1}) + w_n) + \xi_{n+1} \tag{1b}$$

The objective of the direct filter problem is to find the best estimate for the system state parameters $\theta_{n+1}$, given the observational data up to time $t_{n+1}$, usually denoted as $Y_{1:n+1}$, represented by the conditional expectation $\mathbb{E}[\theta_{n+1}|Y_{n+1}]$. In the Bayesian inference framework, the posterior distribution $p(\theta_{n+1}|Y_{1:n+1})$ is estimated at each time step in a two-step prediction-update process. The prediction step combines knowledge from steps $t_n$ with the state model using Chapman-Kolmogorov formula (Eq. 2). The update step incorporates the observations $Y_{n+1}$ with the prior distribution using Bayes' theorem (Eq. 3). When $\mathcal{G}$ and $\mathcal{H}$ are both linear, Eq. 2-3 can be solved analytically using the Kalman formula to give the so-called Kalman filter[1].

$$p(\theta_{n+1}|Y_{1:n}) = \int p(\theta_{n+1}|\theta_n)p(\theta_n|Y_{1:n})d\theta_n \tag{2}$$

$$p(\theta_{n+1}|Y_{1:n+1}) \propto \int p(Y_{n+1}|\theta_{n+1})p(\theta_{n+1}|Y_{1:n})d\theta_{n+1} \tag{3}$$

For general nonlinear, non-Gaussian situations, particle filters are adopted. The main idea of the particle filter method is to use the recursive formulas Eq. 2-3 to generate a cloud of particles whose empirical distribution follows the posterior distribution $p(\theta_{n+1}|Y_{1:n+1})$ so that in the limit of the number of particles, the conditional distribution will converge to the posterior[2]. The framework for the particle filter method is as follows. At time instance $n$, we create a set of M particles $\{x_n^{(m)}\}_{m=1}^M \sim \tilde{p}(\theta_n|Y_{1:n})$ that approximates the conditional pdf $p(\theta_n|Y_{1:n})$ defined by Eq. 4 where $\delta_x$ is the Dirac delta.

$$\tilde{p}(\theta_n|Y_{1:n}) := \frac{1}{M}\sum_{m=1}^{M} \delta_{x_n^{(m)}}(\theta_n) \qquad (4)$$

Then by replacing the prior and posterior with the empirical distributions, the prediction and update steps (Eq. 2-3) become:

$$\tilde{\pi}(\theta_{n+1}|\tilde{Y}_{1:n}) := \frac{1}{M}\sum_{m=1}^{M} \delta_{\tilde{x}_{n+1}^m}(\theta_{n+1}) \qquad (5)$$

$$\tilde{\pi}(\theta_{n+1}|Y_{1:n+1}) := \frac{\sum_{m=1}^{M} p(Y_{n+1}|\tilde{x}_{n+1}^{(m)}) \delta_{\tilde{x}_{n+1}^m}(\theta_{n+1})}{\sum_{m=1}^{M} p(Y_{n+1}|\tilde{x}_{n+1}^{(m)})} \qquad (6)$$

Where $\tilde{\pi}(\theta_{n+1}|\tilde{Y}_{1:n})$ approximates the prior Eq. 2 and $\tilde{\pi}(\theta_{n+1}|Y_{1:n+1})$ approximates the posterior Eq. 3. Finally, we avoid particle degeneracy by introducing a resampling step where we produce more copies of particles with high weights, discard the rest, and normalize the weights. Thus, we obtain $\{x_{n+1}^{(m)}\}_{m=1}^M$ and the "unweighted" empirical distribution given by Eq. 6 which approximates the posterior $p(\theta_{n+1}|Y_{1:n+1})$. Finally, the parameter estimation at instance $t_{n+1}$ is the expectation value of the $\tilde{p}(\theta_{n+1}|Y_{1:n+1})$ (Eq. 7), i.e. the mean value of the particle cloud (Eq. 8) and variance of the particles gives the uncertainty.

$$\tilde{p}(\theta_{n+1}|Y_{1:n+1}) := \frac{1}{M}\sum_{m=1}^{M} \delta_{x_{n+1}}^{(m)}(\theta_{n+1}) \tag{7}$$

$$\hat{\theta} = \frac{1}{M}\sum_{m=1}^{M} \theta_{n+1}^{(m)} \tag{8}$$

Lastly, we constrain the parameter estimates to represent physical values, the rates $k_{ni}$, $k_{gri}$ > 0. This constraint is converted to the equivalent equality constraint $max(-\theta_i, 0) = 0$. We denote the constraints as $G(\theta) = [g_i(\theta)] = 0$ where $g_i$ can represent different equality constraints and we assume that these constraints satisfy a zero-mean Gaussian distribution. We can calculate the likelihood of the constraints with Eq. 9 and incorporate them into Eq. 6 to give a constrained empirical posterior distribution[3].

$$p\left(G(x) = 0 \big| \theta_{n+1}^{(m)}\right) \propto \exp\left(-\frac{1}{2} G(\tilde{\theta}_{n+1}^{(m)})^T \Sigma_c^{-1} G(\tilde{\theta}_{n+1}^{(m)})\right) \tag{9}$$

**Note S2. Synthetic Monolayer Model Parameter Estimation**

Using Eq. 3 in the main text, we generated test data using $k_n = 2.5 \times 10^{-3}$ s$^{-1}$ and $k_{gr} = 5.0 \times 10^{-2}$ s$^{-1}$ with random Gaussian noise and used 500 particles to sequentially estimate the parameters for $t_{i+1}$ at every $t$. We used an initial guess $\theta_0 = (0.0$ s$^{-1}$, $0.0$ s$^{-1})$ with artificial noise variance $\Sigma = (1.0\times10^{-6}$ s$^{-2}$, $1.0\times10^{-3}$ s$^{-2})$ for $k_n$ and $k_{gr}$, respectively. This sequential estimation was repeated 3 times with a different random particle initialization to test the stability and reproducibility of the parameter estimates. Lastly, we visualize the sequentially predicted reflectivity curve $C_r$ for each trial and calculate the average root mean squared error (RMSE) over the 3 trials.

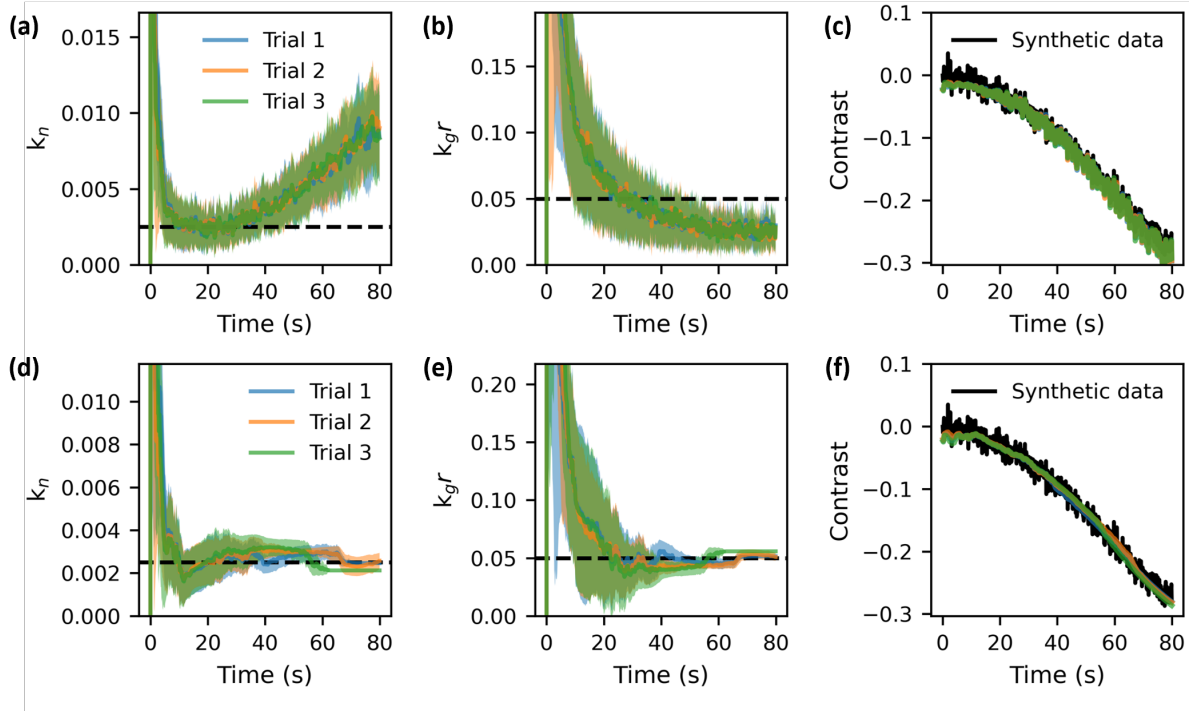

**Figure S1.** Sequential parameter estimation using synthetic data with the monolayer model (Eq. 3, main text) for 3 trials with and without an exploration burn-in period. Each trial is shown as a different color line where the shaded region represents the uncertainty. a) nucleation rate $k_n$ and b) growth rate $k_{gr}$ are estimated for each sequential timestep and c) shows the predicted contrast without implementing a burn-in time for parameter exploration. d-f) show the same estimation but using a 10 s burn-in which more accurately captures the parameter values.

**Figure S1a-b** shows the sequential estimation for $k_n$ and $k_{gr}$, respectively. After ~10 s, $k_n$ converges from the initial guess and begins to oscillate around the true value. The value stays constant up to ~ 30 s and then begins to increase with time. Simultaneously, the $k_{gr}$ estimate converges to a constant value, eventually underestimating the true value. Over all 3 trials, the estimation behavior is consistent. In **Figure S1c**, we take the estimated parameters at every time point and calculate the predicted reflected contrast $C_r$ for each trial and plot them with the experimental data. Although the DF did not converge to the true parameter values, the curve is reproduced remarkably well, even at very early times, with an average RMSE = 0.010. If the goal in the experiment is to sequentially reconstruct the reflectivity curve, this is acceptable. In our case,

however, we are interested in accurately estimating the model parameters quickly in order to understand the growth kinetics in real-time.

The divergence from the true parameters is caused by an increase in the number of particles given 0 weight as time goes on. Methods for dealing with this issue include particle inflation[4] or rejuvenation[5]. Here, we choose to use a so-called 'burn-in' period to aid in parameter convergence. The artificial noise $\epsilon_n$ is analogous to a learning rate and as the number of 0 weighted particles increases, the artificial noise can lead to divergence. To implement a burn-in period, we allow exploration during the early times and then set $\epsilon_n = 0$ after some prescribed time. Based on the time of convergence for $k_n$ in **Figure S1a**, we select a burn in time to be when the contrast $C_r$ has reached ~ 0.01 which represents 3% monolayer surface coverage. **Figure S1d-e** shows the results of implementing a 10 s burn-in time. We observe that after the burn-in period, the parameter estimates now tend to converge to near the true value and the reduced estimation noise leads to a smoother predicted $C_r$ curve. In this case, the DF converges near the true values and accurately reproduces the $C_r$ curve with an average RMSE of 0.013. We also tested this same burn-in period for 10x slower growth rates, using $k_n$ = 2.5×10$^{-4}$ s$^{-1}$ and $k_{gr}$ = 5.0×10$^{-3}$ s$^{-1}$ with similar success.